\documentclass{isprs} 
\usepackage{subfigure}
\usepackage{setspace}
\usepackage{listings}
\usepackage[hyphens]{xurl}
\usepackage{geometry} 
\usepackage{epstopdf}
\usepackage{xcolor}
\usepackage[labelsep=period]{caption}  
\usepackage[british]{babel} 
\usepackage[hang]{footmisc}

\usepackage{hyperref}
\hypersetup{
    colorlinks=true,
    linkcolor=blue,
    urlcolor=blue,
    citecolor=blue
    }
\colorlet{punct}{red!60!black}
\definecolor{background}{HTML}{EEEEEE}
\definecolor{delim}{RGB}{20,105,176}
\colorlet{numb}{magenta!60!black}
\lstdefinelanguage{json}{
    basicstyle=\footnotesize
\ttfamily,
    numbers=none,
    numberstyle=\scriptsize,
    stepnumber=1,
    numbersep=8pt,
    showstringspaces=false,
    breaklines=true,
    backgroundcolor=\color{background},
    literate=
     *{0}{{{\color{numb}0}}}{1}
      {1}{{{\color{numb}1}}}{1}
      {2}{{{\color{numb}2}}}{1}
      {3}{{{\color{numb}3}}}{1}
      {4}{{{\color{numb}4}}}{1}
      {5}{{{\color{numb}5}}}{1}
      {6}{{{\color{numb}6}}}{1}
      {7}{{{\color{numb}7}}}{1}
      {8}{{{\color{numb}8}}}{1}
      {9}{{{\color{numb}9}}}{1}
      {:}{{{\color{punct}{:}}}}{1}
      {,}{{{\color{punct}{,}}}}{1}
      {\{}{{{\color{delim}{\{}}}}{1}
      {\}}{{{\color{delim}{\}}}}}{1}
      {[}{{{\color{delim}{[}}}}{1}
      {]}{{{\color{delim}{]}}}}{1},
}

\begin{document}

\title{Integrating multimedia documents in 3D city
models for a better understanding of territories}
\date{30/04/2022}

\author{C. Gautier\textsuperscript{1,}\thanks{Corresponding author}, J. Delanoy\textsuperscript{2}, G. Gesquière\textsuperscript{3}}
\address{
 	\textsuperscript{1 }Univ Lyon, UCBL, CNRS, INSA Lyon, LIRIS, UMR5205, F-69622 Villeurbanne, France - corentin.gautier@universite-lyon.fr\\
 	\textsuperscript{2 }Univ Lyon, INSA Lyon, CNRS, UCBL, LIRIS, UMR5205, F-69621 Villeurbanne, France  - johanna.delanoy@insa-lyon.fr\\
 	\textsuperscript{3 }Univ Lyon, Univ Lyon 2, CNRS, INSA Lyon, UCBL, LIRIS, UMR5205, F-69676 Bron, France  - gilles.gesquiere@univ-lyon2.fr\\
}


  \commission{IV }{} 
\workinggroup{ IV/9 } 
\icwg{}   

\abstract{
	
	Digital 3D representations of urban areas, through their growing availability, are a helpful tool to better understand a territory.
	However, they lack contextual information about, for example, the history or functionality of buildings.
	On another side, multimedia documents like images, videos or texts usually contain such information.
	Crossing these two types of data can therefore help in the analysis and understanding of the organization of our cities.
	This could also be used to develop document search based on spatial navigation, instead of the classical textual query.
	In this paper, we propose four approaches to integrate multimedia documents in a 3D urban scene, allowing to contextualize the scene with any type of media. We combine these integration approaches with user guidance modes that allows to guide the user through the consumption of these media and support its understanding of the territory.
	We demonstrate the usefulness of these techniques in the context of different projects within  the Lyon area (France).
	The use of multimedia documents integrated into a digital tour allows, for example, the iconic buildings to be contextualised or to understand the evolution of a territory through time.

}

\maketitle

\section{Introduction}\label{sec:Introduction}
Digital 3D representations of urban areas are becoming widely available. Multiple tools exist to visualize them such as Google Earth\footnote{https://www.google.com/intl/fr/earth/} or Cesium\footnote{https://cesium.com/}. By allowing to wander into virtual cities, these tools allow a better understanding of a territory. However, most of them are limited to a 3D representation of the buildings, without providing more information about their functionality, their evolution or other elements of context. Aside from 3D representations, multimedia (e.g. images, texts or videos) also plays an important role in the understanding of the urban landscape. Archival images, videos on the history of a district, job descriptions of certain industries or interviews in a district can provide important additional information about a territory. While these media have the potential to enrich virtual 3D cities, the link between those multimedia and digital 3D representations is rarely made.

Online mapping systems (e.g. Google Maps) allow to wander in a 2D or 3D view and pick information through labels or pins that simulate points of interest. By linking multimedia to a spatial position, users can find information more easily in a desired area. This can be seen as another form of search, rather than formulating a query on a web browser. As such, these tools lack ways to fully integrate the media into a 3D view of the city, and usually lack of explanatory media on the organisation of a district or a whole city, on its history, etc.

\begin{figure*}[th!]
	\begin{center}
			\includegraphics[width=1.5\columnwidth]{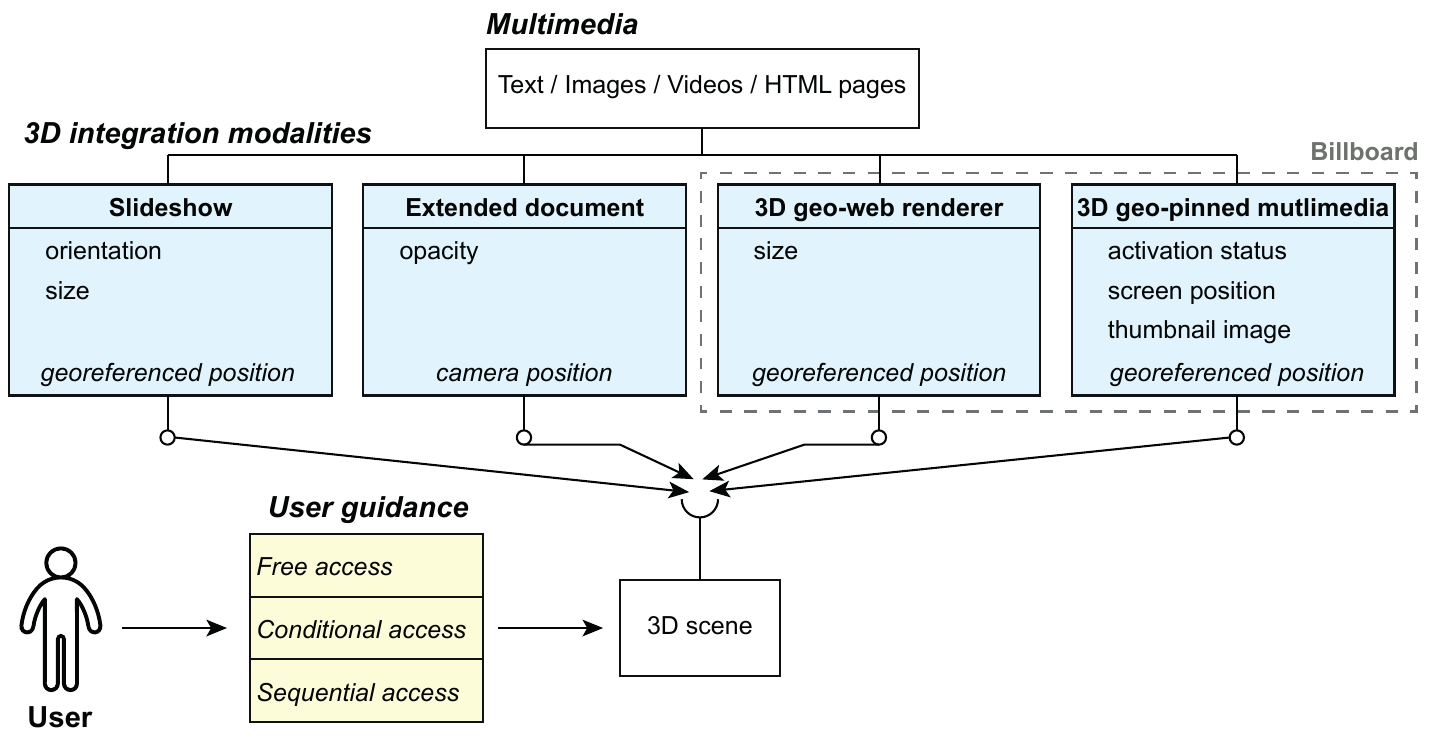}
		\caption{Architecture diagram of our proposed model. The multimedia can be integrated in the 3D scene using one of the multimedia integration approaches (in blue). Each approach is characterized by attributes that allow to parameterize the integration. The media are linked to the scene by either a georeferenced position or a camera position (shown in italic). The user can consume the integrated media in the scene through different guidance modes (in yellow).}
	\label{fig:overview}
	\end{center}
\end{figure*}

We propose a model that allows to combine a 3D view of the city in which the user can navigate with a collection of multimedia that provides additional information about the underlying 3D scene. We present new ways to integrate each multimedia document in the 3D scene as well as ways to guide the user through a collection of different documents. More precisely, we introduce four modalities of multimedia integration in a scene (shown in blue in Figure \ref{fig:overview}) that are presented in Section \ref{sec:integration}:
\begin{itemize}
	\item The \textbf{3D geo-pinned multimedia} uses the principle of pins that target a point of interest and are linked to a multimedia document. Our extended pins gives an overview of the linked document.
	\item The \textbf{3D geo-web renderer} shows documents at given positions in the scene, always facing the camera such that the user can get the information from any place in the scene.
	\item The \textbf{extended document} superimposes images on a 3D model and complete them with textual data. The extended document is linked to a precise camera position.
	\item The \textbf{slideshow} allows to make a presentation in a  view by placing documents in a 3D space.
\end{itemize}

A large number of documents linked in a scene do not necessarily guarantees a better understanding of a territory, as users might need to be guided through them in order to understand them. We thus present ways to guide the user through its consumption of documents in the scene. Our three user guidance modes range from a fully sequential order of consumption, to a free mode where the user can wander in the scene and pick the documents as he wishes. These modes and the way they can can be mixed with the multimedia integration modalities are presented in Section \ref{sec:consume}.

The combinations of the integration modalities and user guidance modes open the door to a large variety of choices in the way to use multimedia documents in combination with 3D virtual cities. 
We demonstrate the potentialities of several of these combinations in the context of multiple projects developed in the city of Lyon (France) in Section \ref{sec:implementations}. Such projects include a web-documentary that aims to deconstruct preconceived ideas about a territory or presentations about the future of a neighbourhood using a tangible model.

\section{Related works}\label{sec:Related works}

Many tools exist to navigate in a virtual urban environment \cite{bleisch_quantitative_2015,boutsi_interactive_2019,jaillot_integrating_2021}.
Thanks to the navigation in the scene, users can visualize a city or any 3D geometry under different angles. This can greatly help to better understand an urban space by virtually navigating in it.

Other tools, such as web-documentaries, also allow to walk in an urban space and better understand it without making use of a 3D scene.
These tools can target a wider audience as they are web-based and do not require any specific hardware.
Their 2D representation of the digital territory is enhanced by additional information such as video ("Correspondances" \footnote{ https://correspondances.tv5monde.com/ }) or images  
("Balade au merlan" \footnote{http://unebaladeaumerlan.fr/}).
However, these tools remain limited in their interaction because of the 2D representation of the world that they carry. 
We would like to take up this spatial contextualization of documents to the third dimension and integrate such documents in 3D walking tours.

Enhancing 3D scenes with multimedia documents have been used to show the evolution of a city through time \cite{jaillot_integrating_2021,chagnaud_visualization_2016}.
Such documents can bring information about the state of a city at a certain date through images of buildings at different times or their chronology.
These works have introduced several methods to visualize historical photo documents in a 3D scene, allowing to show the urban past but also the future planning of cities.
However, they lack the possibility to provide  additional information on these pictures to help reach a better understanding. They also do not allow to use any other type of media, such as videos of web pages, or to interact with them.

Other methods have tried to contextualise a digital model with quantitative data \cite{bleisch_quantitative_2015} or text and images \cite{boutsi_interactive_2019} in a scene. 
This gives another perspective on the data and a better understanding of a territory by linking information to a geospatial position.
These methods are good examples on how a 3D representation and multimedia or data can complement each other.
However, they only integrate one type of content and do not give an overview of the different documents present in the scene, making the search for a specific document or data harder.
We want to provide a new method of document search through a digital representation of a territory by allowing an explicit overview of the different multimedia available.

Rather than images or data, web pages can also be integrated in a 3D context to contextualise a 3D scene \cite{wijnants_web-mediated_2015}.
This allows giving the user access to more content, such as external websites, while not losing the focus on the digital representation of a territory.
The interest of integrating the web page directly into the 3D scene is also to support the relationship between the target and its related annotation objects \cite{seo_webized_2015}.
While we follow the same goal, we aim to provide a model that allow to integrate any type of multimedia in the scene and to bring information on a specific geometries like building or district
\cite{samuel_representation_2016}.

Multimedia documents can also have interesting metadata like their title, source, publication date, key content, tags. \cite{gan_document_2014} gives an overview of the existing techniques used to visualize a multitude of documents both in 2D and 3D environment. However, they do not target geographic applications and the integration in 3D urban scenes.

While integrating multimedia document in the 3D digital representation of the city can greatly improve the understanding of such area, the user can quickly be lost in this multitude of information and not know where to start searching.
Some works have shown the interest in guiding the user by making him follow a logical sequence of information.
It also helps to engage the user and allow him to be more effective in its reading of the information \cite{othman_engaging_2011}.
In our work, we combine methods to integrate multimedia into 3D urban scenes with ways to interact with a collection of documents, providing different levels of guidance to the user.

Existing methods of integrating multimedia do not satisfy our need to manage multiple types of document or to provide interaction with them.
We provide several ways to integrate such multimedia in the 3D scene, depending on the type of such documents but also on their goal and content. These integration modes can be combined with user guidance modes that help reach  better user experience and understanding of territory.

\section{Multimedia integration approaches}\label{sec:integration}

Our objective is to integrate different types of multimedia documents into a 3D urban scene in order to make the representation more informative and interactive. The integrated documents can thus be videos (eventually 360 videos) as well as images, textual documents or web pages, as shown in Figure \ref{fig:overview}.
They are all georeferenced and have a position in space.
Although this position is a natural link with the 3D scene describing the city, integrating the media visually in a simple interface such that the user can mentally connect them to the 3D representation of the city is not easy.

We propose four methods to integrate such documents in the 3D scene, explained thereafter and shown in blue in Figure \ref{fig:overview}. The choice of one of them depends on the type of media but also on their meaning or their number.
This integration makes it possible to have all the information on a district on the same web interface, easing the understanding of such areas. But it can also help to find documents or information by wandering in the digital representation of the city and browsing the multimedia that are available.
Each of these integration modalities can be used  in the 3D scene with different user guidance modes (free access, conditional and sequential) that will be described in Section \ref{sec:consume} and are shown in yellow in Figure \ref{fig:overview}.

The first two approaches (\textbf{3D geo-pinned multimedia} and \textbf{3D geo-web renderer}) integrate a media into a 3D scene such that it is linked to a precise spatial location and always shown facing the camera (similar to the billboard technology). The \textbf{extended document} is linked to a camera position and superimposes the media to the 3D scene, while the \textbf{slideshow} projects a series of media onto a 3D plane in the scene. All of these methods aim to be integrated in a web-based library for 3D visualization of geographical data.

\subsection{Billboard inspired approaches}
These approaches consist in an arrangement of thumbnails directly shown in the 3D scene and targeting a precise spatial point. 
The thumbnail is always facing the camera such the user can access the information from any position in the scene.
By directly showing the multimedia content into the 3D scene, these thumbnails allow to contextualize buildings and places without taking the user out of its 3D wandering. 
The two approaches differ in the way the document is displayed through the thumbnail: either directly or summarized in an interactive pin.
In either cases, the integration is characterized by a georeferenced position (GPS coordinates).

\begin{figure}[ht!]
	\begin{center}
			\includegraphics[trim={0 0 150px 0},clip,width=0.9\columnwidth]{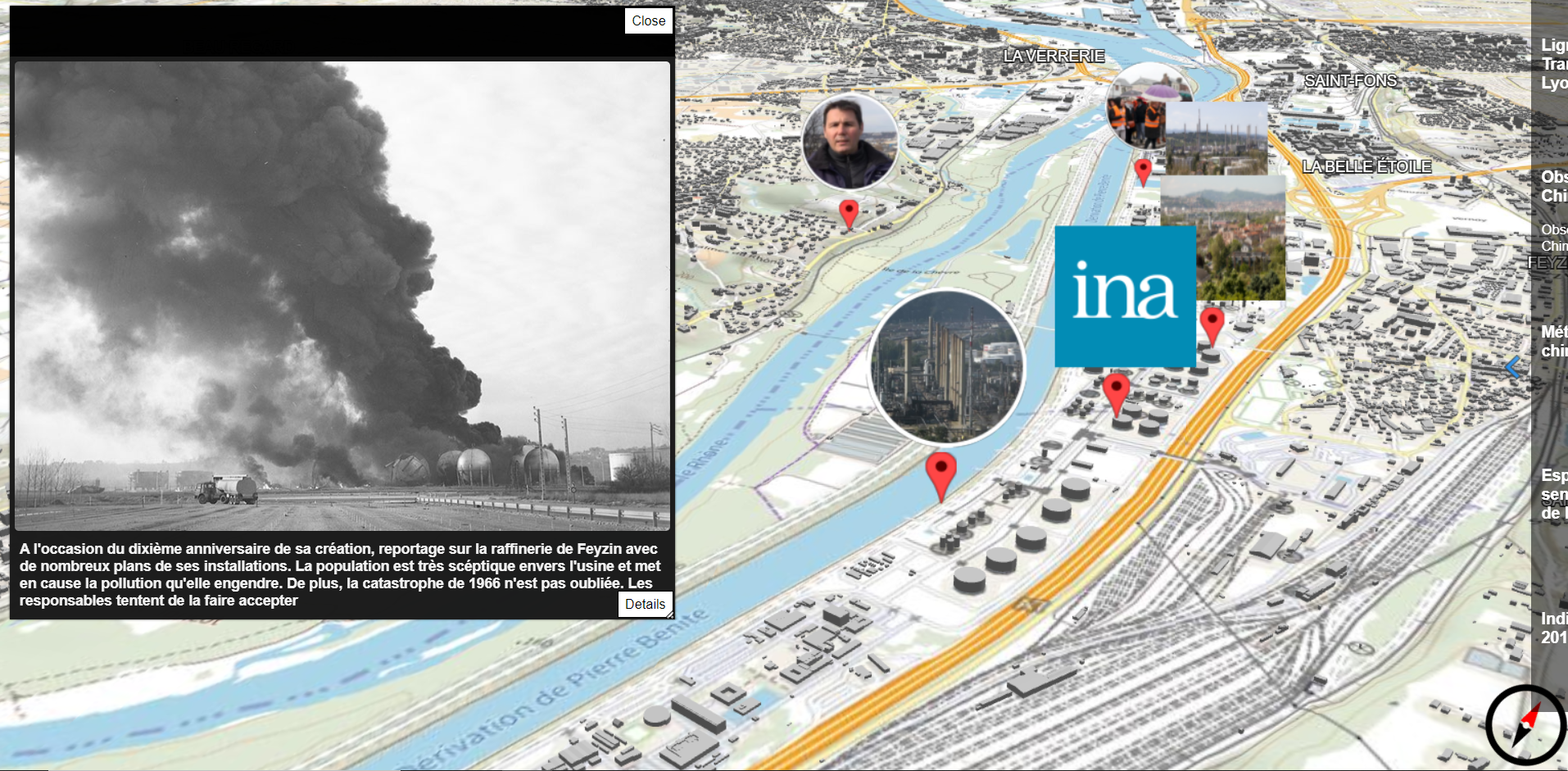}
		\caption{An example of the 3D geo-pinned multimedia approach. The pins show the available multimedia on the 3D scene while the selected document is shown on 2D at the left of the screen.}
	\label{fig:pinned}
	\end{center}
\end{figure}

The \textbf{3D geo-pinned multimedia} is based on the principle of pins that target a point of interest. 
Such kind of pins, in the shape of an inverted drop
\footnote{https://en.wikipedia.org/wiki/Google\_Maps\_pin},
is one of the most effective way of highlighting points of interest on a map and is well understood by most users.
We enrich this traditional approach by adding a thumbnail above it that gives a visual feedback on the nature and content of the multimedia. 
This pin is meant to be interactive: when selected, the linked document 
appears in 2D on the screen in front of the 3D scene (see example in  Figure \ref{fig:pinned}). The pin and its thumbnail are always facing the camera while the multimedia document is displayed as a 2D element that will not move with the camera.  This approach allow to give the user a global vision of all the contents available in a scene. 
The user can then pick from this data and decide to get more detailed information about a place or building pointed out by the pins. 

The interactivity of the pin can be deactivated, allowing to control the content the user can access at a given moment, as it will be explained in Section \ref{sec:consume}.
In this case, the user can still see the pin and its thumbnail, thus knowing that there is a document accessible, but will not be able to consume the linked document. 
In addition to its spatial position, the 3D geo-pinned multimedia is thus also characterized by its activation status, the position where to display the document on the screen and its thumbnail image (see Figure \ref{fig:overview}).

\begin{figure}[ht!]
	\begin{center}
			\includegraphics[width=1\columnwidth]{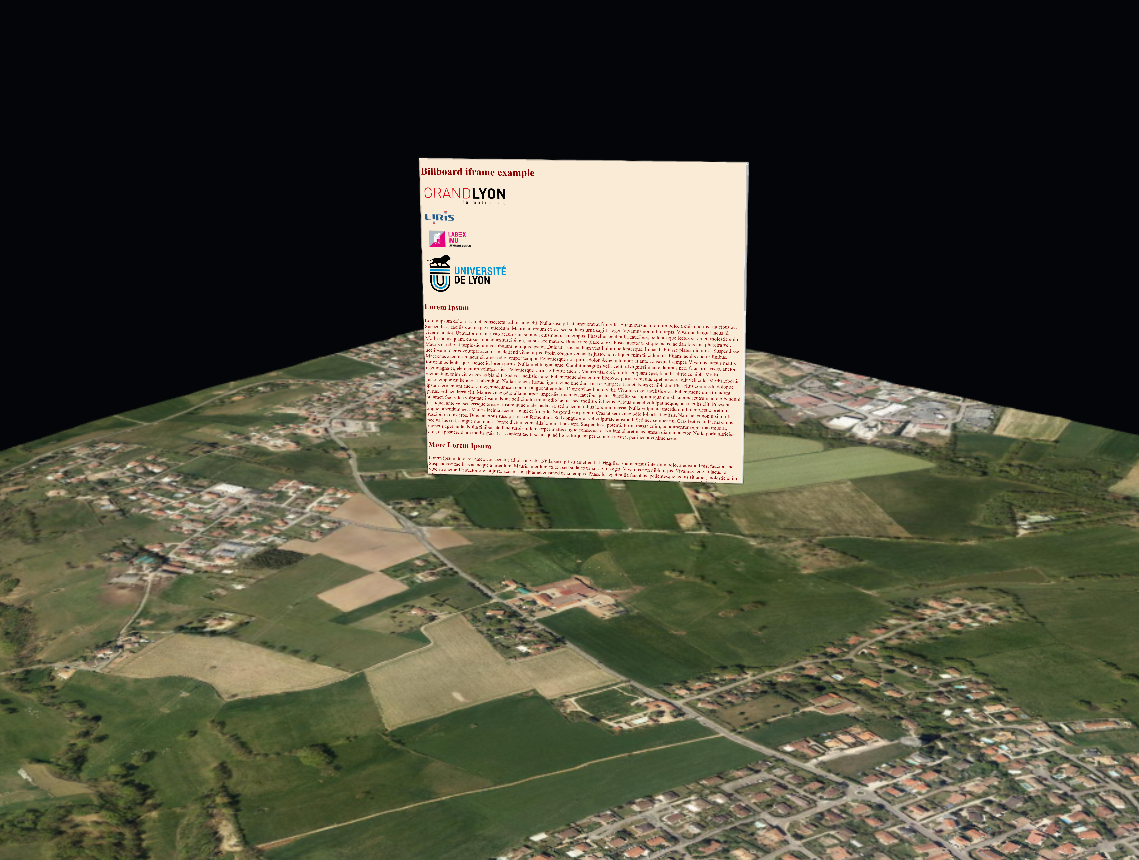}
		\caption{An example of the 3D geo-web renderer approach. An html page is shown over the 3D scene.}
	\label{fig:web}
	\end{center}
\end{figure}

The \textbf{3D geo-web renderer} also points to a precise location in the 3D scene but the linked document will be directly shown to the user when he approaches it. Again, the document will be displayed always facing the camera so that the user can see it from any position in the scene. This way, the user can continue walking around in the 3D scene and mentally put this document in context. An example of the integration of a web page with this approach is shown in Figure \ref{fig:web}. The size the media appears in the 3D scene is an additional parameter in this integration method.
 
These two billboard approaches are similar in the way they are integrated into a scene, they both target a precise location in space and allow to quickly see which content is accessible while moving into the scene.
However, the 3D geo-pinned multimedia summarizes the information into small pins instead of directly showing it, allowing to display more document without cluttering the space. It could thus be a preferred solution when a large number of documents need to be shown in a small space.
Both are naturally good approaches to integrate web pages as they can point to an url, allowing the user to directly browse a web page from within the 3D scene.

\subsection{Extended Document}

The \textbf{extended document} approach displays a multimedia document in a 2D context on top of a 3D scene. 
The integration of the document in the 3D scene is done by modifying the position and orientation of the camera, rather than positioning the document in the 3D space (see Figure \ref{fig:inspectorWindow}). 
Typical multimedia for this approach would be   images or videos that give a vision of the city in another time, past or future. The traditional method to show the evolution of a city is to compare two images placed side by side. The extended document allows to visualize the evolution of a place in a different way, by superimposing the image and the digital 3D model, helping to better understand the changes.
In order to access different documents in the same scene, 
the user has to choose the documents he is interested in in a list. When selected, the camera will move to the right position and overlay the document on top of the 3D scene. In order to keep the link with the 3D scene, the camera will move smoothly toward the target position, allowing the user to locate more easily its position in 3D.
Documents in this approach are thus characterized by the position and orientation of the camera, as well as the opacity of the element to be overlaid.

The extended document is composed of three distinct windows:
\begin{itemize} 
	\item The \textbf{navigator windows} (Figure \ref{fig:inspectorWindow}, left-most window) which allows to navigate between the different documents with their titles and reference dates. This window also has a filter system for a better navigation among the documents .
	\item The \textbf{inspector windows} (Figure \ref{fig:inspectorWindow}, right-most window) gives a preview of the document that will be superimposed in the 3D scene, as well as all additional information about it such as its description, source, and publication date. It allows the user to have a first understanding of the document and verify that it is the media that he wants to consume.
	\item The \textbf{visualization windows} (Figure \ref{fig:inspectorWindow}, middle window) is activated when the user decides to engage the visualization in the inspector window. It will smoothly move the camera to the correct position and superimpose the document on the 3D view. It also allows to change the opacity of the document in order to compare it with the underlying 3D.
	
	\begin{figure}[ht!]
	\begin{center}
			\includegraphics[width=1.0\columnwidth]{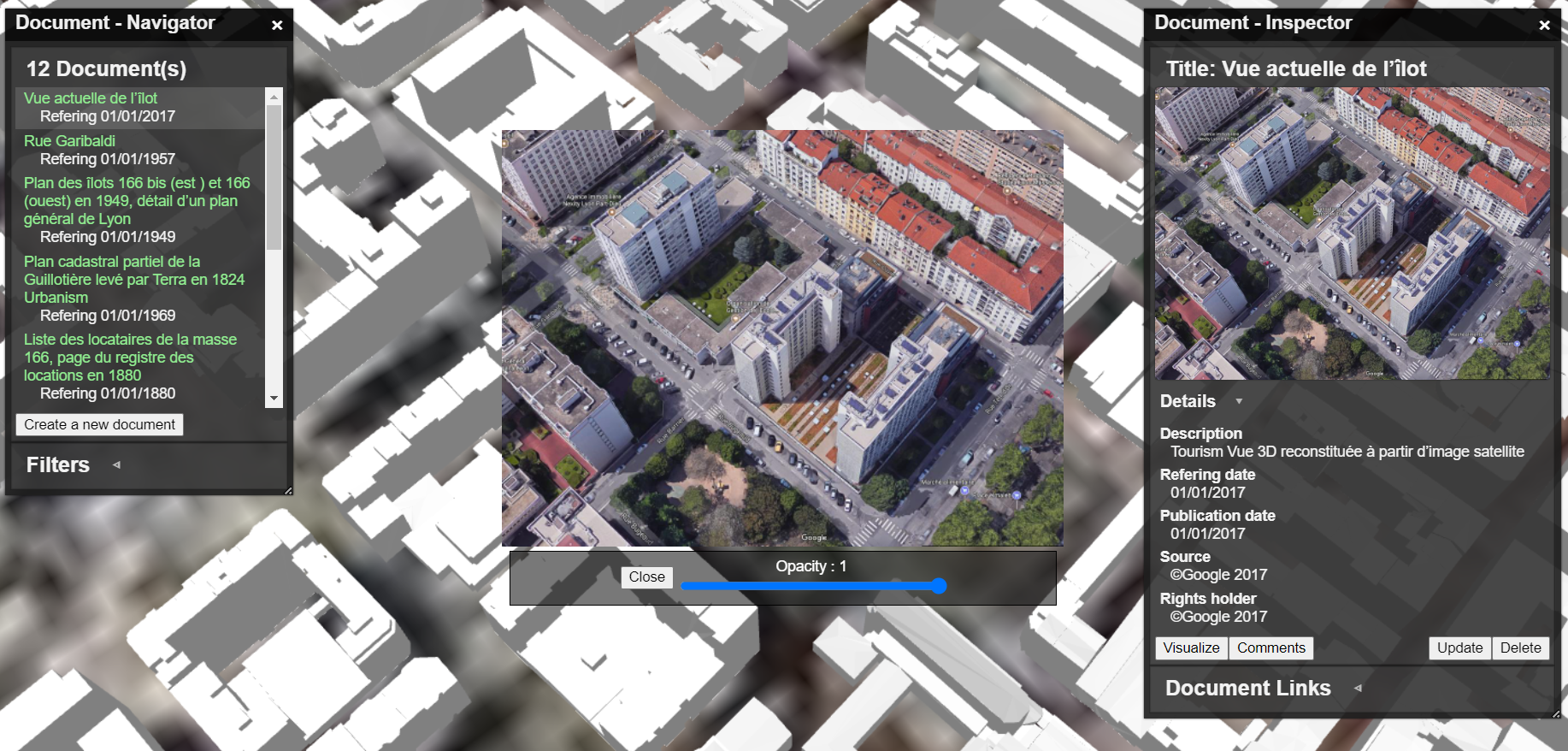}
		\caption{An example extended document with its three windows: the navigator window showing the list of available documents (left); the inspector window showing the metadata of the selected document (right); the visualization window that superimposes the document on the 3D scene (middle).}
	\label{fig:inspectorWindow}
	\end{center}
\end{figure}
\end{itemize}

This approach will be preferably used with images, or animated images, as it allows to directly compare it with the 3D view of the scene. It can help users to realize the evolution of a district or a building, or observe it under another format.

\subsection{Slideshow}

\begin{figure}[ht!]
	\begin{center}
			\includegraphics[width=1\columnwidth]{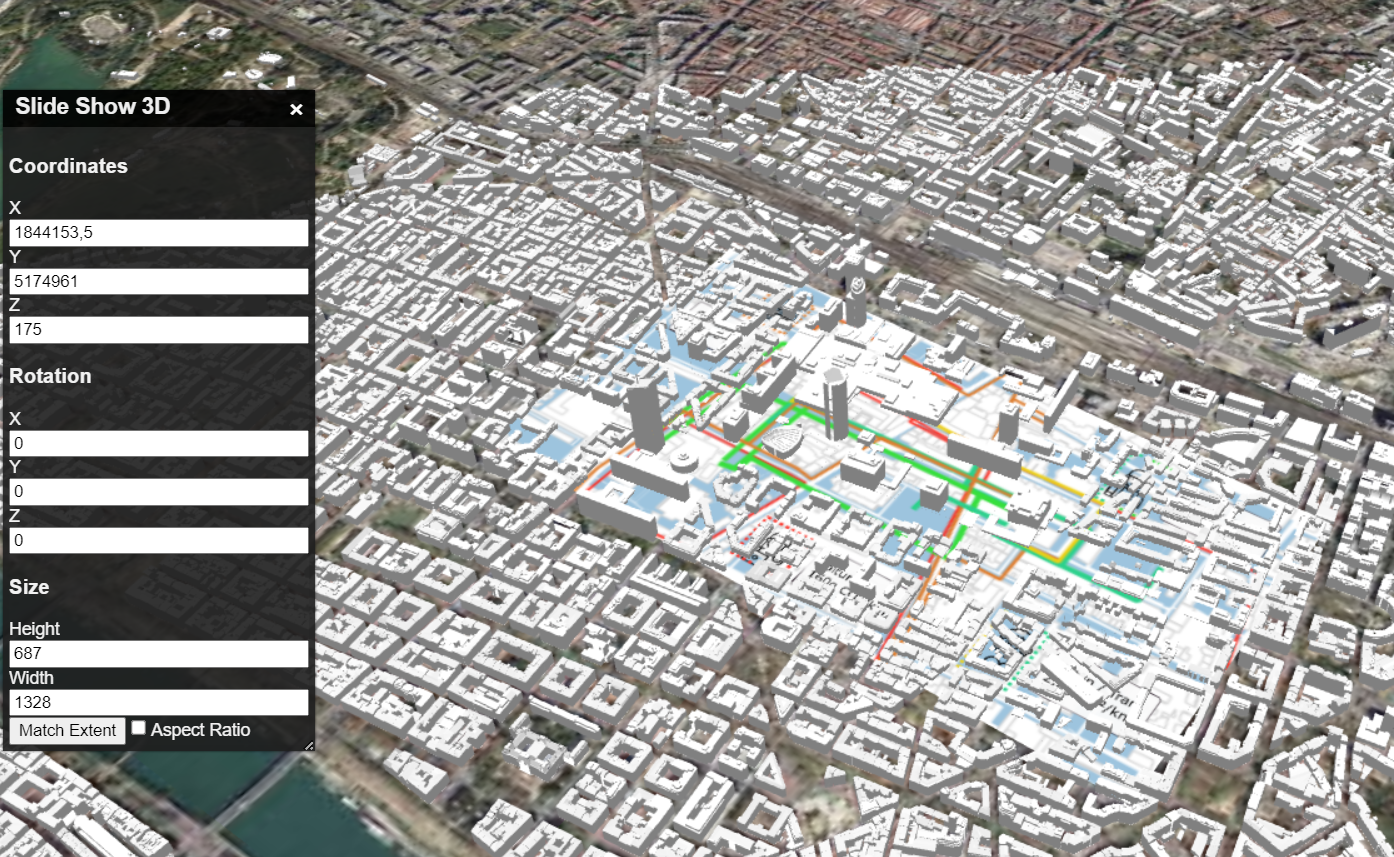}
		\caption{An example of the slideshow.
		The configuration window (on the left) allows to set its parameters. One image of the slideshow is shown on a ground plane in the 3D scene.}
	\label{fig:slideshow}
	\end{center}
\end{figure}

The slideshow is anchored at a 3D position in the 3D scene and can display the documents on a 3D plane (see Figure \ref{fig:slideshow}).
The slideshow is made to integrate a series of media like images or animated images: it simulates a presentation in the 3D world in which one can scroll through different documents, located at the same place, in order to illustrate problems or ideas on a district.
This plane is traditionally parallel to the ground but can be also set vertically to contextualize face buildings.  The documents will thus not follow the camera as the user move into the scene. The slideshow is characterized by its center position (georeferenced position), its size and its orientation.
This method overlays the 3D model of buildings on top of the documents, that can be images or videos of data layers on a district, or different states of a territory. For example, one could drop the image of the transport network of a city and see how it fits with the morphology of that city.
The user can easily drag and drop different images or videos into the scene, allowing to quickly illustrate different ideas, but also to work in team and experiment different solutions and illustrations. A geographic mask is made available to prepare data directly in dedicated  GIS information systems tools.

\section{User guidance}\label{sec:consume}

Our different multimedia integration approaches allow to integrate a large variety of multimedia
in a scene, from an archive image to a web page.
When becoming too numerous in one scene, 
the user could get lost among them thus breaking the goal of our approach to bring a better understanding of a territory.
We thus design several ways to consume a collection of documents, by possibly guiding the user through their consumption:

\begin{itemize}
    \item \textbf{Sequential access}: The documents are provided in a given order. As in a guided tour, this allows to build a narrative and guide very precisely the user through the documents. This mode limits the order of freedom of the user, it can only act on the time to spend on each document, but helps to build a better understanding of the documents. 

    \item \textbf{Conditional access}: Going through some documents might be needed to access others, but the order is not completely constrained. Several documents are available at one moment in which the user can chose freely, but viewing some or all of them is needed to access the next ones.

    \item \textbf{Free access}: the user can consume the different multimedia in the scene as he wishes.
\end{itemize}

Most of our integration approaches can be used with any of these user experiences, although some are more naturally suited for some specific user guidance. The only exception is the slideshow, that is essentially made for sequential access. It allows a linear succession of images or videos on a 2D map and can be used in the case of a presentation to contextualise a district.

The extended document, because it has smooth transitions between documents and is not anchored at a specific position in the 3D scene, is well suited to be used in a sequential approach. It can be used to build a \textit{guided tour} as a list of extended document where each of them with be contextualized in the tour with additional text. It can, for example, consist of a sequence of documents that show the evolution of a district 
 with photos from different eras.

Both the 3D geo-pinned multimedia and the 3D geo-web renderer can be used naturally with any of the guidance.
However the 3D geo-pinned multimedia is well suited for conditional access thanks to its activation status. It allows the user to see all the contents that is available in the scene through the pins, but the activation of some of the documents can be conditioned to the viewing of other ones.

\section{Implementations }\label{sec:implementations}
We experimented our different approaches on various use cases around the city of Lyon (France). Each of these use cases mobilizes one of our multimedia integration approaches, combined with a user guidance mode, allowing a wide variety of user experiences.
These implementations are implemented within the UD-Viz library\footnote{https://github.com/VCityTeam/UD-SV},
a JavaScript library allowing to visualize, analyse and interact with urban data.

\subsection{Chemistry Valley: 3D geo-pinned multimedia }

We demonstrate the use to the 3D geo-pinned multimedia in the context of a web-documentary, developed as part of a collaboration between a training center for the Chemistry industry Interfora\footnote{https://www.interfora-ifaip.fr/} and the Lyon metropolis.
This web documentary aims to make the citizens rediscover the area of the Chemistry Valley (South of Lyon, France) through a 3D digital wandering where the user can interact with documents arranged in the 3D scene. 

\begin{figure}[ht!]
	\begin{center}
			\includegraphics[width=1.0\columnwidth]{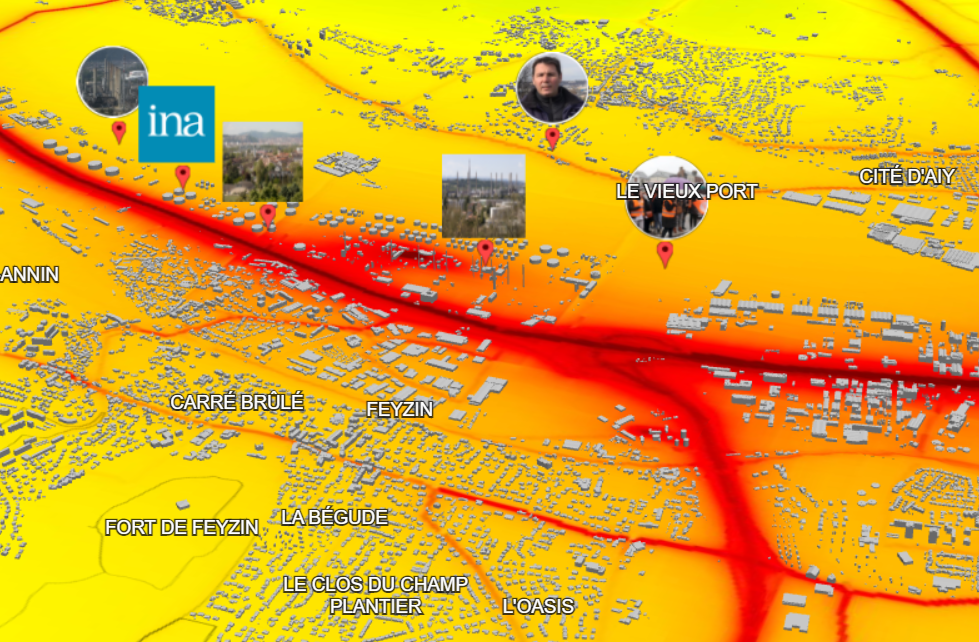}
		\caption{The 3D geo-pinned multimedia modality used in the context of the Chemistry valley. The pins show an overview of the different media available in the scene.}
		\label{fig:chemistry_valley}
	\end{center}
\end{figure}

A large number of media, from different types were to be set in the scene. As examples:
\begin{itemize} 
	\item A photographic observatory \footnote{https://www.caue69.fr/1/page/10622/page} of the Chemistry valley, consisting of a set of pictures of the chemistry valley from different angles.
	\item Additional layers as for instance bus lines, the atmospheric index, etc. can be used thanks to numerous data available via WFS (Web Feature Service) and WMS (Web Map Service) in the open data data.grandLyon.com\footnote{https://data.grandlyon.com/}.
	\item Jobs descriptions located in the chemistry area. These pdf files provide information on workers in the area.
	\item Videos of stakeholder interviews in the Chemistry valley on the theme of employment and training. 
\end{itemize}

The 3D geo-pinned multimedia is the best choice in this case since it allows to easily navigate between a large number of media, without cluttering the scene.
Additionally, some of these media are clustered in different themes which need to be seen in a given order, naturally leading to using a conditional access mode.
  As a reminder, in the  3D geo-pinned multimedia modality, documents can be naturally deactivated.
Their pin and thumbnail are still shown on the 3D scene but the user can not interact with them to access the document. In order to better guide the user towards the accessible content, we symbolize the deactivation of the media by a lock image superimposed to the thumbnail.By guiding the user through the different themes, this visualization helps the user to better understand the whole issue.
An example of a visualization of the Chemistry valley in this project can be seen in Figure \ref{fig:chemistry_valley}: the atmospheric index is used as a color map on the ground of the 3D model while several pins show the available content.

\begin{figure}[hbtp]
    
    \begin{lstlisting}[language=json,frame=single]
"episode-1-data":{
 "content-1" :{
  "lock" : false, // document activation
  "position" :{ // coordinates in the scene
   "x" : "1843554.77",
   "y" : "5165405.73",
   "z" : "220"
   },
   "imgUnlock":"./assets/img/Observatoire/cheminee.jpg",
   "imgLock":"./assets/img/Observatoire/cheminee.jpg",
   "text": "Vallee de la chimie - Observatoire photographique",
   "src":"https://umap.openstreetmap.fr/fr/map/vallee-de-la-chimie"
 }
    \end{lstlisting}
	\caption{An example of a JSON configuration file used for the 3D geo-pinned multimedia.}
	\label{code.1}
\end{figure}

Each interactive element in this scene can be configured through a JSON file that contains the geospatial position, the display thumbnail, its activation status, the text content and the document link. An example of such configuration file can be seen in Figure \ref{code.1}.

The integration of multimedia in the 3D scene makes it possible to show the different professions that exist in the Chemistry valley, and help the users to know more about the Chemistry industry in this territory.
To further this understanding, we have also integrated urban data into the scene by using geoservices\footnote{https://geoservices.github.io/}, spatial web services that make geodata available in a structured form.
Finally, in order to show the valley differently, we have integrated WMS (Web map service) and WFS (Web feature service)  which allow the display of georeferenced geometries such as building cadastres.

Thanks to the 3D geo-pinned multimedia method, 
we have brought another form of information search through the wandering in the 3D scene.
The user can pick up interviews with the stakeholders of this territory and get information on where they are located and which company they belong to before even starting to watch the video.
This makes it easier for him to choose the content that interests him.

\subsection{Flying campus: 3D geo-web renderer}

\begin{figure}[ht!]
	\begin{center}
			\includegraphics[width=1.0\columnwidth]{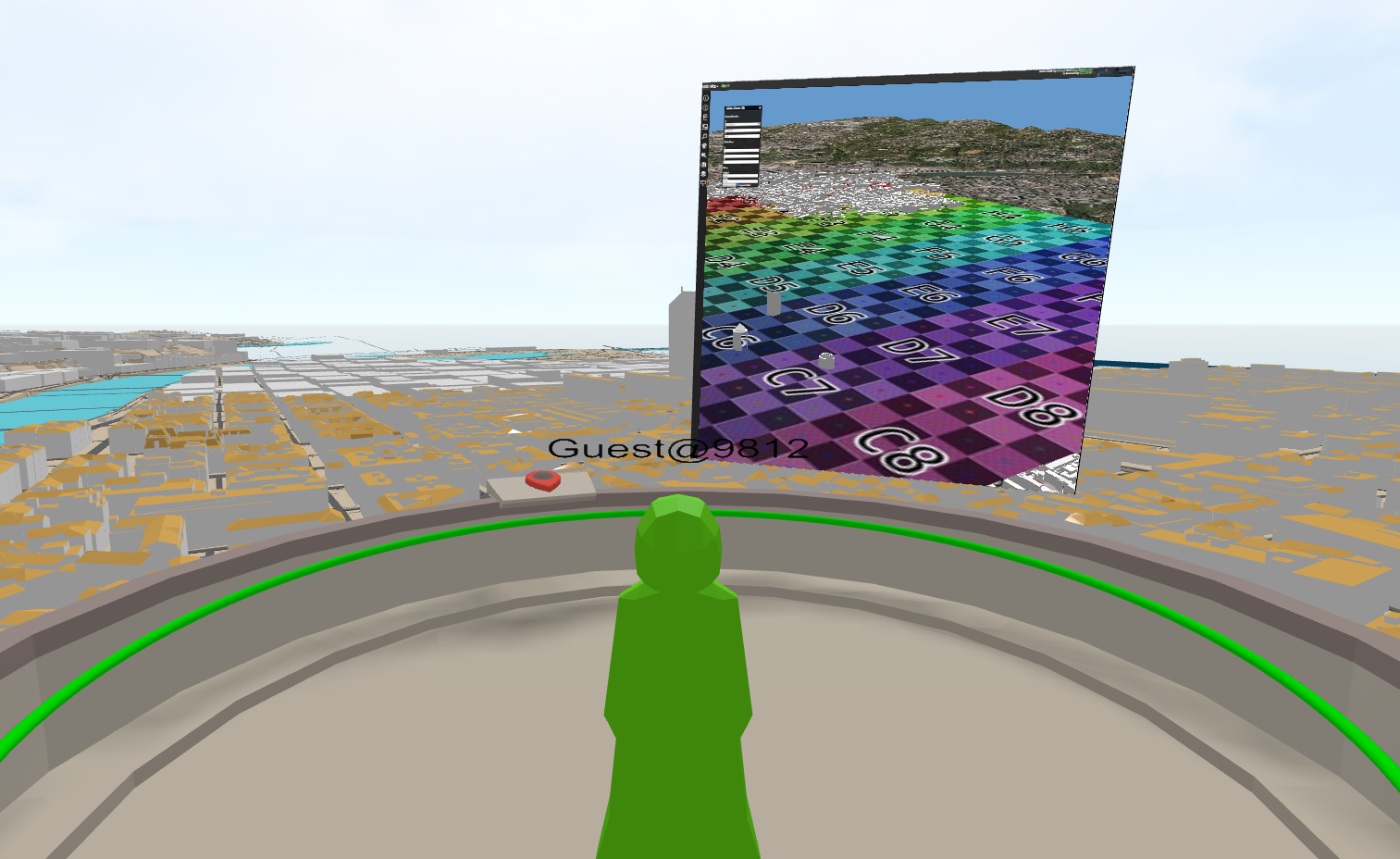}
		\caption{An example of the use of the 3D geo-web renderer in "Flying campus", a virtual gathering place. The user can still observe the 3D scene while the integrated multimedia is shown to the user.}	\label{fig:flyingcampus}
	\end{center}
\end{figure}

We use the 3D geo-web renderer method in the context of the project "IMUV/Flying campus"\footnote{https://demo.liris.cnrs.fr/vcity/flyingcampus/}. The aim of this project is to create a virtual space in which photographic exhibitions, conferences, or other events can be organised. It is an interactive space, floating above the city, where a variety of elements can be arranged and where several users can gather. Elements to be displayed can be of any types, including web pages. 
The use of the 3D geo-web renderer allow to integrate any type of media into the 3D space, such that the user can still be in immersion while consuming the content. One of the use case is a conference room in which a 3D geo-web renderer is placed to broadcast a video, the users can gather in the room and freely interact with the video. Figure \ref{fig:flyingcampus} shows a user observing the city in 3D while looking at a map showing the organisation of the city.

We implemented the 3D geo-web renderer using the CSS3Drenderer technology (a web rendering method from the javascript library ThreeJS\footnote{https://threejs.org/}) that allows to include web pages, with which the user can interact, in the 3D scene.
Our implementation consists in an overlay of two types of rendering: a CSS3Drenderer  which shows the document to be displayed and the WebGLRenderer which is the 3D scene rendering of the walkthrough \cite{seo_webized_2015}.
We make a part of the 3D scene rendering transparent and render the CSS3DRenderer context behind the WebGLRenderer context. This allow the content of the document in the CSS3DRenderer to appear in the 3D scene.

By placing these billboards in the 3D scene of the flying campus, the user does not loose focus and has all the information in one web application. If the document to integrate is a web page, the user can freely interact with it as if it was facing the window of a web browser.

In this context, the user is free to consume the content of these multimedia as he wishes and is not limited by an order to follow (free access). 

Flying campus demonstrates the value of this multimedia integration as a way to virtually share multimedia experience and build participative experiences.

\subsection{Historical guided tour: Extended document}

\begin{figure}[ht!]
	\begin{center}
			\includegraphics[width=1.0\columnwidth]{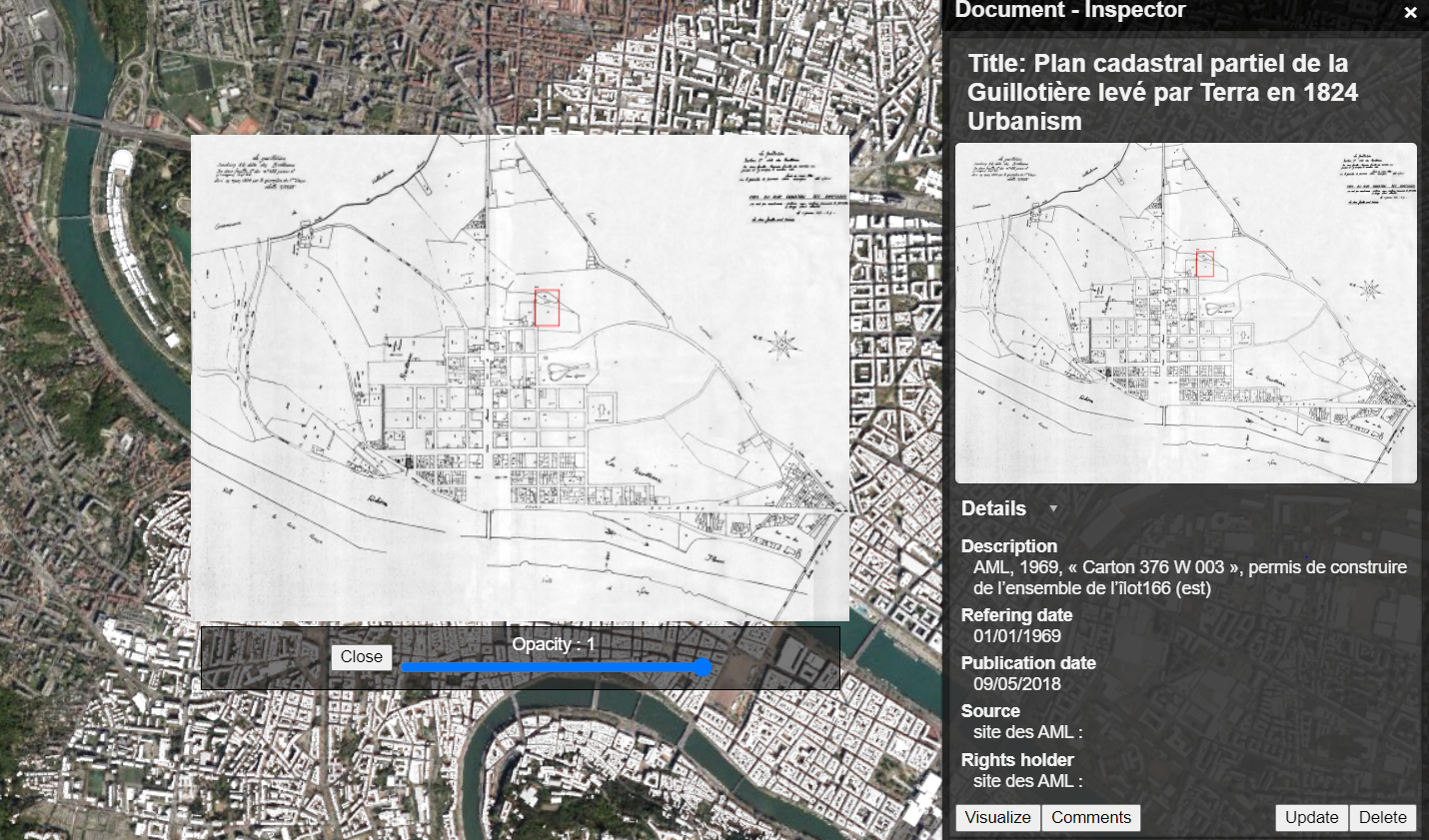}
		\caption{An historical  cadastral plan is shown superimposed on top of the 3D scene by using the extended document modality.}	\label{fig:extendeddoc}
	\end{center}
\end{figure}

In this project, we use the extended document as a way to show the evolution of a district over time. Our goal was to integrate a collection of archival photos from 1760 to 2017 in the 3D scene. Each picture shows how the district looked like at a specific date along with additional information. By allowing to superimpose the picture on top of the 3D scene, at its correct position, the extended document is naturally suited for this kind of use cases. By playing on the opacity slider, 
the user can more easily observe the morphological evolution of a building or a district between two periods. An example showing the historical organisation of a district with some additional information is shown in Figure \ref{fig:extendeddoc}

To bring more coherence between these documents, we used the sequential access mode as a user guidance. We thus developed a \textbf{guided tour} in which the user can follow the chronological evolution of the district and better understand how it has evolved through the ages.

This use case allowed us to provide information about the evolution of a district thanks to archive images and the integration in the 3D scene. 

\begin{figure*}
    \centering
    \includegraphics[width=0.20\textwidth, angle=90]{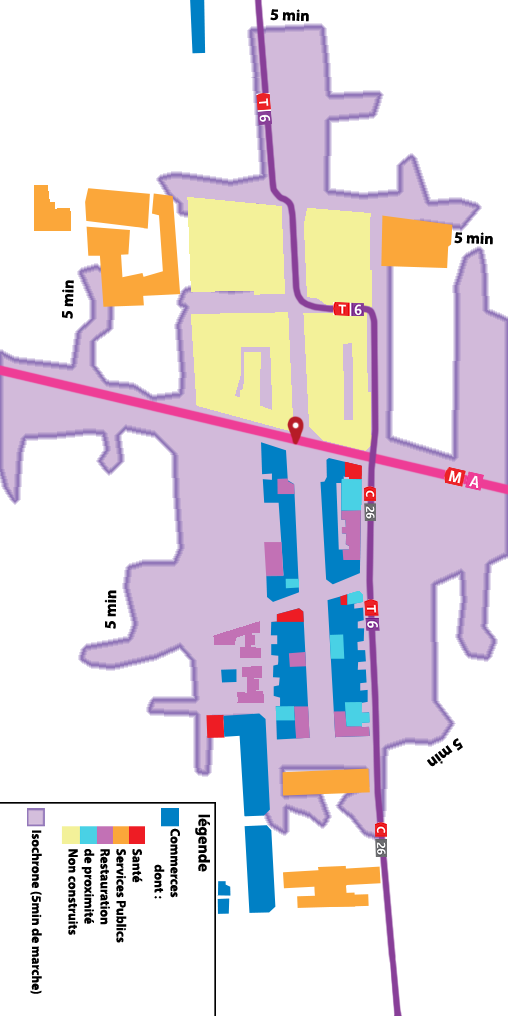}
    \includegraphics[width=0.20\textwidth, angle=90 ]{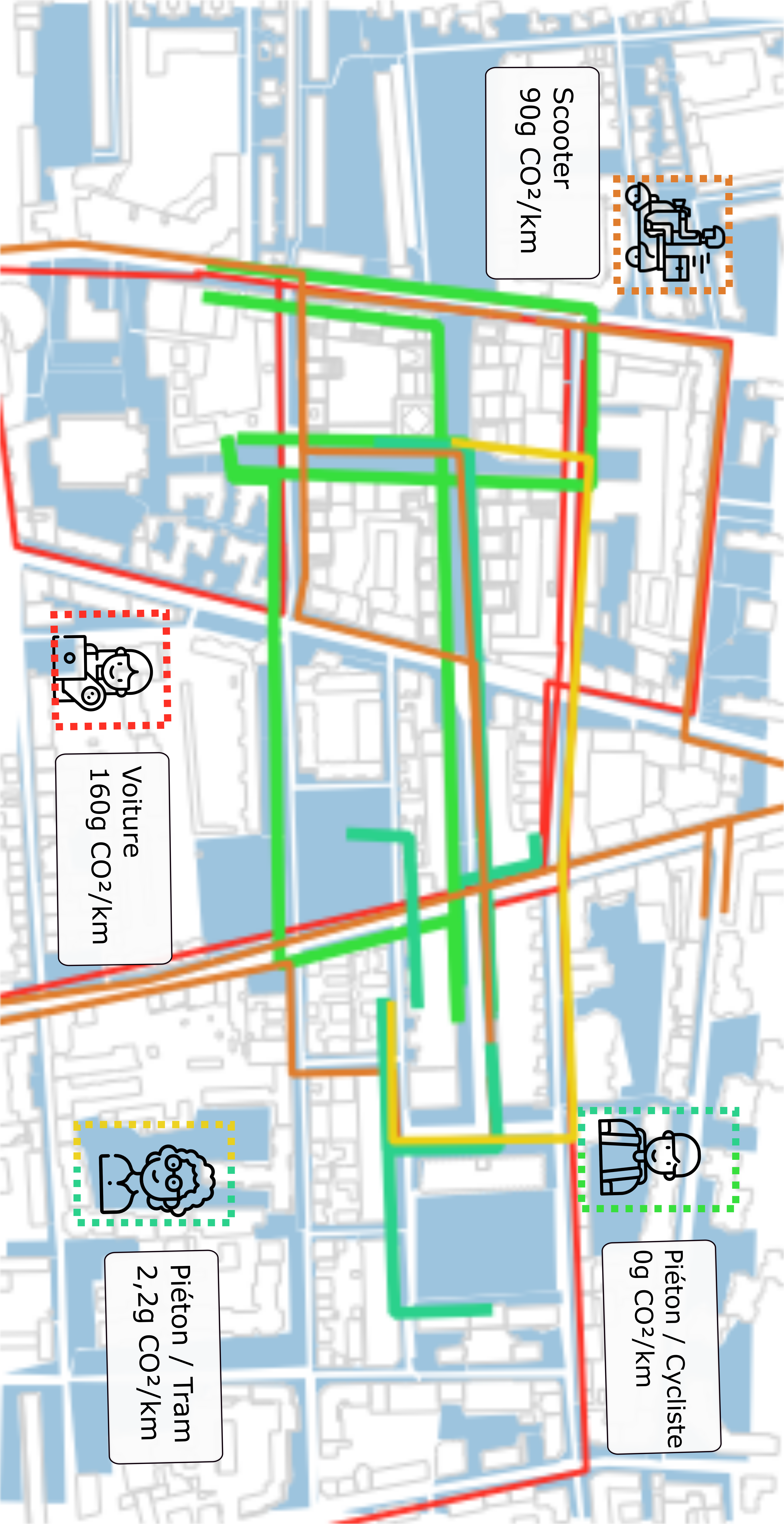}
    \includegraphics[width=0.20\textwidth, angle=90]{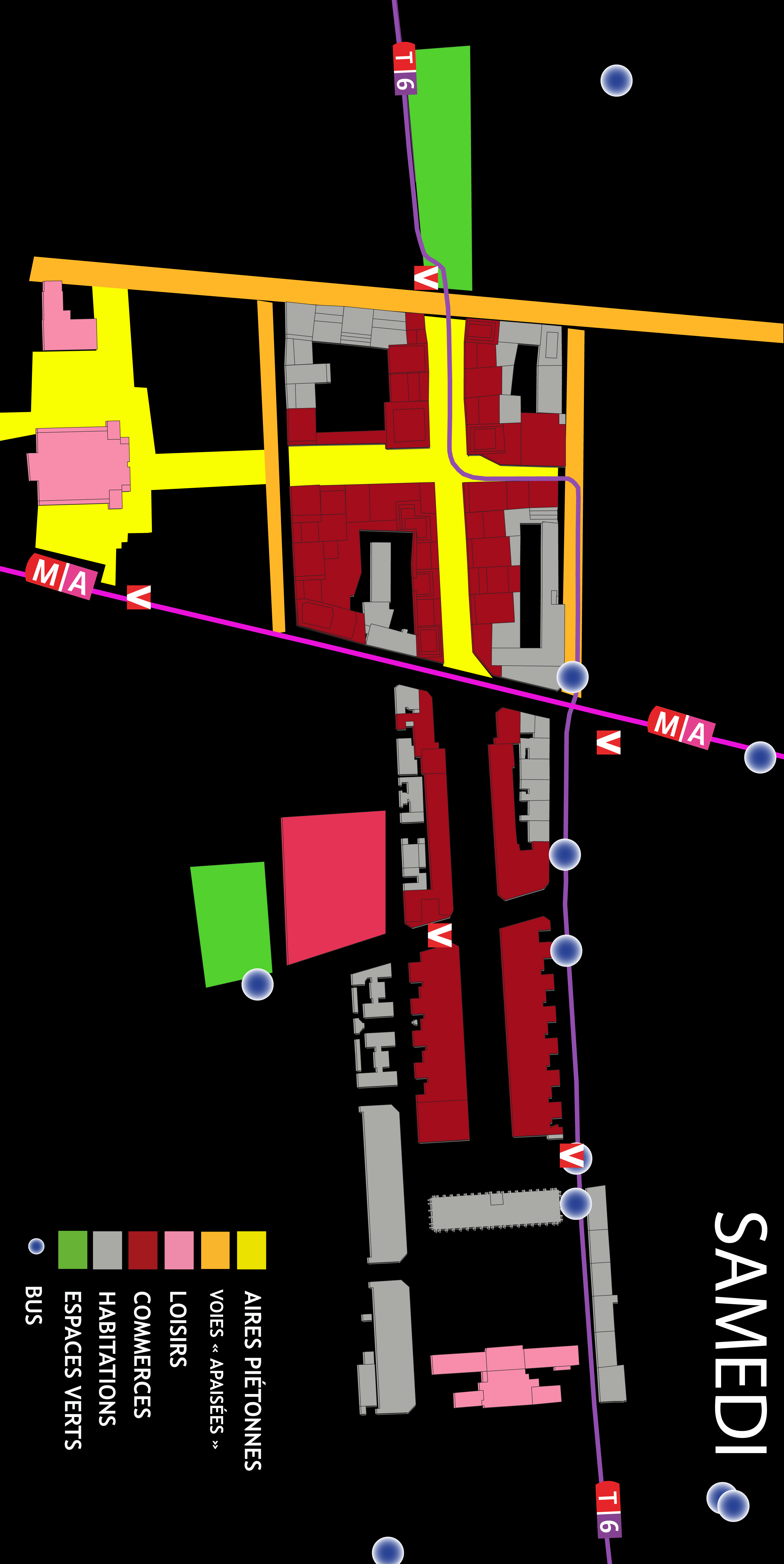}
    \caption{Three examples of image layers created by students for the slideshow. These layers can be used one after another in a presentation.}    \label{fig:gratteciellayers}
\end{figure*}

\subsection{Gratte-Ciel project: Slideshow}

We use the slideshow approach in the context of an event organized during the  Anthropocene week\footnote{https://ecoleanthropocene.universite-lyon.fr/a-quoi-revent-les-maquettes--247850.kjsp?RH=1633680335198}.
Several multidisciplinary students (Geomatician, urban planner, information-
communication) gathered on the issue of using a tangible or digital model as a technological tool for the intelligibility of territories.
The students produced images or videos as an information layer over a district of Lyon, that they could contextualize with the 3D model of the same district. Three examples of such layers are shown in Figure \ref{fig:gratteciellayers}. The images were produced such that they fit the underlying 3D geometry and can thus be superimposed with it. The slideshow method allows to see the geometry of building on top of these layers, allowing to better understand the interaction between those buildings and the data represented in the different layers.

The students could drop their vizualisations onto the 3D map and animate a presentation by going through the different maps.
They have proposed still image of geographic data or animated simulation with multi-agent model. 

\begin{figure}[ht!]
	\begin{center}
			\includegraphics[width=1.0\columnwidth]{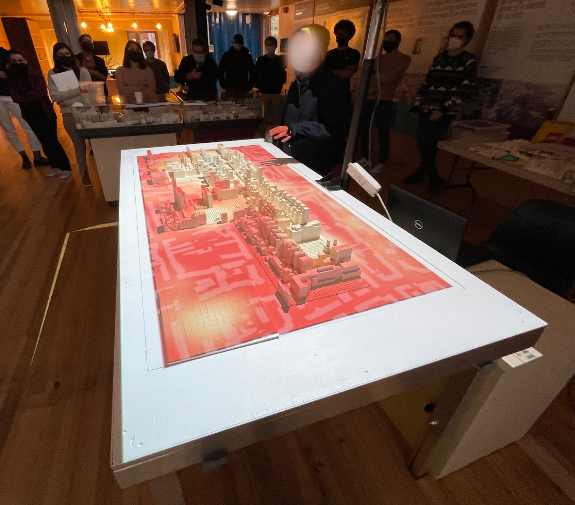}
		\caption{One image of the slideshow projected onto the tangible model.}
		\label{fig:grattecielmodel}
	\end{center}
\end{figure}

The aim of this project was also to show how tangible models could help in understanding urban projects. In addition to the 3D scene, we also provided the students with a tangible model of the neighbourhood that was built using white Legos. The white color allows to easily project information into the model. Rather than only seeing their maps in the digital 3D scene, students could project them onto the tangible model.  Figure \ref{fig:grattecielmodel} shows an example projection on the model. We calibrated a video-projector such that the 3D scene in UD-Viz was matching with the tangible model, and we projected the slideshow module on the model. Using the slideshow on a tangible model allows to more easily understand and discuss about the interaction between the building and other data.

In these examples, we could observe the interest in adding an image to a 3D scene and having a way to present information linked to the 3D world.
The mix of 3D modelling of buildings and videos or map backgrounds allowed to better understand the problematic of the district and the possibilities of the new urban project.

\subsection{Building an home made application}
All the described components are delivered as open source content available at  \textcolor{blue}{https://github.com/VCityTeam/UD-Viz}.
Each of the demonstration project can be tested using the following links:

\begin{itemize}

    \item Chemistry valley project using the 3D geo-pinned multimedia:
    \url{https://github.com/VCityTeam/UD-Demo-TIGA-Webdoc-ChemistryValley}
    \item Flying campus using the 3D geo-web renderer:
    \url{https://github.com/VCityTeam/UD-Viz/blob/master/examples/Billboard.html}
    \item Historical guided tour using the extended document approach:
    \url{https://github.com/VCityTeam/UD-Demo-VCity-Spatial-multimedia-db-Lyon}
    The docker can be found at
    \url{https://github.com/VCityTeam/UD-Demo-VCity-Spatial-multimedia-db-Lyon}
    \item Gratte-ciel project with the slideshow approach: \url{https://github.com/VCityTeam/UD-Demo-Anthropocene-GratteCiel-}
    The docker can be found at \url{https://github.com/VCityTeam/UD-Demo-Anthropocene-GratteCiel-docker}
     
\end{itemize}

Some demos are provided as dockers to ease their deployment. All the source code is accessible such that the different integration methods can be exploited in other contexts.

\section{Conclusion}\label{sec:Conclusion}

Multimedia documents play an important role in understanding the urban landscape.
We have presented several ways to integrate multimedia in a 3D scene in order to bring a better understanding of a territory. We also introduced different user guidance modes, going from a fully guided sequence up to a free consumption of media, than can be combined with our multimedia integration modalities. 
Different use-cases around real project allowed us to demonstrate the variety of user experiences that can be created with these approaches.
The addition of multimedia in the urban digital model brings another dimension to it, and greatly improve the way user can understand and discover a territory.

We currently demonstrated use cases making use of a unique integration approach for all the media in the scene, but the different approaches can be combined in the experience.
Each integration method can contextualise the digital model in different ways and provide more information to the user. These approaches are widgets of the UD-Viz library and can therefore be easily integrated into the same scene. That opens interesting avenues for future research on how these different modalities can interact with each other and can greatly extend the possibilities of our model. 

While our integration approaches can cover various scenario and multimedia types, it might be complicated to find the right integration method for a given use case. It is also not clear yet what are the exact effects on the user experience for each of these modes. Additionally, while our user guidance approach helps the user to find its way in presence of a lot of content, a filtering or grouping system would also prove useful. Being able to assign themes or time period to the different contents would allow the user to filter the different media, but also to group them in the scene and develop new modalities to interact with them. For example, linking the multimedia with a time period would add a fourth dimension to the scene, allowing to travel through time in the scene. The user could then observe the evolution of a district or a building by navigating through documents of different ages.

\section*{ACKNOWLEDGEMENTS}\label{ACKNOWLEDGEMENTS}

The authors would like to thank the TIGA project, funded by La Banque des Territoires, led by the metropolis of Lyon and the members of the Vcity projects of LIRIS lab \footnote{https://liris.cnrs.fr/} for the fruitful discussion and help in this work.

{
	\begin{spacing}{1.17}
		\normalsize
		\bibliography{main_commented.bib}
	\end{spacing}
}
\end{document}